\documentclass[preprintnumbers,nofootinbib,showpacs]{revtex4}

\usepackage{graphicx}
\usepackage{latexsym}
\usepackage[english]{babel}
\usepackage{amsmath}
\usepackage{amsthm}
\usepackage{epsfig}
\usepackage{color}
\usepackage{amssymb}

\begin{document}

\title{\Large Dynamics driven by the Trace Anomaly in FLRW Universes}

\preprint{ITP-UU-09/52, SPIN-09/43}


\author{Jurjen F. Koksma}
\email[]{J.F.Koksma@uu.nl} \affiliation{Institute for Theoretical
Physics (ITP) \& Spinoza Institute, Utrecht University, Postbus
80195, 3508 TD Utrecht, The Netherlands}

\begin{abstract}
By means of a semiclassical analysis we show that the trace
anomaly does not affect the cosmological constant. We calculate
the evolution of the Hubble parameter in quasi de Sitter
spacetime, where the Hubble parameter varies slowly in time, and
in FLRW spacetimes. We show dynamically that a Universe consisting
of matter with a constant equation of state, a cosmological
constant and the quantum trace anomaly evolves either to the
classical de Sitter attractor or to a quantum trace anomaly driven
one. There is no dynamical effect that influences the effective
value of the cosmological constant.
\end{abstract}

\maketitle

\section{Introduction}
\label{Introduction}

This paper aims at summarising the main results of
\cite{Koksma:2008jn}, as presented at the ``Invisible Universe
Conference'' in Paris (2009). We show that the cosmological
constant problem is not solved by the trace anomaly.

In semiclassical gravity, one treats all matter fields as quantum
fields living on a classical curved background. The fields, i.e.:
their classical (background) expectation values and their
fluctuations, cause the background to curve according to
Einstein's field equations. The question how these fluctuations
affect the background geometry, or in particular the cosmological
constant, is a line of research known as quantum backreaction. We
need to combine classical general relativity with knowledge of
quantum field theory in curved spacetimes \cite{Birrell:1982ix}.

We supplement the classical Einstein-Hilbert action with the trace
anomaly or conformal anomaly which quantum field theories are
known to exhibit \cite{Capper:1974ic, Deser:1976yx, Brown:1976wc,
Dowker:1976zf, Tsao:1977tj, Duff:1977ay, Brown:1977pq,
Birrell:1982ix}. Suppose a classical action is invariant under
conformal transformations of the metric, then the resulting
stress-energy tensor is traceless:
\begin{equation}\label{intro1}
T^{\mu}_{\phantom{\mu}\mu} = 0 \,.
\end{equation}
As an explicit example, consider a conformally coupled scalar
field. In quantum field theory the stress-tensor is promoted to an
operator. A careful renormalisation procedure renders its
expectation value $\langle
\hat{T}^{\mu}_{\phantom{\mu}\nu}\rangle$ finite. However,
inevitably, the renormalisation procedure results in general in a
non-vanishing trace of the renormalised stress-energy tensor:
\begin{equation}\label{intro2}
\langle \hat{T}^{\mu}_{\phantom{\mu}\mu}\rangle \neq 0 \,.
\end{equation}
Classical conformal invariance cannot be preserved at the quantum
level. In short, this is the trace anomaly. An important question
that immediately arises is how this non-zero trace affects the
evolution of the Universe through the Einstein field equations. In
particular, one could wonder whether it influences the effective
value of the cosmological constant.

It has been argued (see \cite{Antoniadis:2006wq} and references
therein) that the trace anomaly could potentially provide us with
a dynamical explanation of the cosmological constant problem. In
brief, the line of reasoning is as follows. Firstly, a new
conformal degree of freedom is introduced as
$g_{\mu\nu}(x)=\exp[2\sigma(x)]\overline{g}_{\mu\nu}(x)$.
Furthermore, the trace anomaly stems from a non-local effective
action that generates the conformal anomaly by variation with
respect to the metric \cite{Riegert:1984kt}. The authors of
\cite{Antoniadis:2006wq} then argue that the new conformal field
should dynamically screen the cosmological constant, thus solving
the cosmological constant problem.

The proposal advocated in \cite{Antoniadis:2006wq} is very
appealing. Before studying the effect of this new conformal degree
of freedom, we feel that firstly a proper complete analysis of the
dynamics resulting from the effective action of the trace anomaly
should be performed. This is what we pursue in this contribution.

We argue in favour of a semiclassical approach to examining the
connection between the cosmological constant and the trace
anomaly: we take expectation values of inhomogeneous quantum
fluctuations with respect to a certain state to study its effect
on the background spacetime. Phase transitions aside, quantum
fluctuations affect the background \textit{homogenously}.
Moreover, in accordance with the Cosmological Principle,
inhomogeneous fluctuations of the metric tensor and in particular
of the conformal part of the metric tensor are observed to be
small at the largest scales, comparable to the Hubble radius (and
expected to be small also in the early Universe). We do not
consider a new, conformal degree of freedom. Hence, we do
certainly not exclude any possible effect this (inhomogeneous)
conformal degree of freedom might have on the cosmological
constant. However, it is plausible that in order to address the
link between the cosmological constant and the trace anomaly, a
semiclassical analysis suffices.

A final note is in order regarding an application of the conformal
anomaly: trace anomaly induced inflation. In the absence of a
cosmological constant, the trace anomaly could provide us with an
effective cosmological constant \cite{Starobinsky:1980te,
Hawking:2000bb, Pelinson:2002ef,Shapiro:2002nz, Shapiro:2003gm,
Pelinson:2003gn, Shapiro:2008sf}. If one includes a cosmological
constant, the theory of anomaly induced inflation is plagued by
instabilities, which we will also come to address. Improving on
e.g. \cite{Pelinson:2002ef, Brevik:2006nh}, we incorporate matter
with a constant equation of state in the Einstein field equations.

\section{The Dynamics driven by the Trace Anomaly}

The trace anomaly or the conformal anomaly in four dimensions is
in general curved spacetimes given by \cite{Capper:1974ic,
Birrell:1982ix, Antoniadis:2006wq}:
\begin{equation}\label{TraceAnomaly1}
T_{\mathrm{Q}}\equiv \left\langle
\hat{T}_{\phantom{\mu}\mu}^{\mu}\right\rangle = b F + b'
\left(E-\frac{2}{3}\Box R\right) + b'' \Box R \,,
\end{equation}
where:
\begin{subequations}
\label{TraceAnomaly2}
\begin{eqnarray}
E &\equiv& \mbox{}^{*}R_{\mu\nu\kappa\lambda}
\mbox{}^{*}R^{\mu\nu\kappa\lambda} = R_{\mu\nu\kappa\lambda}
R^{\mu\nu\kappa\lambda}- 4 R_{\mu\nu}
R^{\mu\nu} + R^{2} \label{TraceAnomaly2a} \\
F &\equiv& C_{\mu\nu\kappa\lambda} C^{\mu\nu\kappa\lambda} =
R_{\mu\nu\kappa\lambda} R^{\mu\nu\kappa\lambda}- 2 R_{\mu\nu}
R^{\mu\nu} + \frac{1}{3} R^{2} \label{TraceAnomaly2b} \,,
\end{eqnarray}
\end{subequations}
where as usual $R_{\mu\nu\kappa\lambda}$ is the Riemann curvature
tensor, $\mbox{}^{*}R_{\mu\nu\kappa\lambda}
=\varepsilon_{\mu\nu\alpha\beta}
R^{\alpha\beta}_{\phantom{\alpha\beta}\kappa\lambda}/2$ its dual,
$C_{\mu\nu\kappa\lambda}$ the Weyl tensor and $R_{\mu\nu}$ and $R$
the Ricci tensor and scalar, respectively. The Gauss-Bonnet
invariant is denoted by $E$. The general expression for the trace
anomaly can also contain additional contributions if the massless
conformal field is coupled to other long range gauge fields (see
e.g. \cite{Birrell:1982ix}). Finally, the parameters $b$, $b'$ and
$b''$ appearing in (\ref{TraceAnomaly1}), dimensionless quantities
multiplied by $\hbar$, are determined by the (matter) degrees of
freedom in a theory as follows:
\begin{subequations}
\label{TraceAnomaly3}
\begin{eqnarray}
b &=& \frac{1}{120(4\pi)^{2}}\left( N_{S}+6N_{F}+ 12 N_{V}\right)
 \label{TraceAnomaly3a} \\
b' &=& -\frac{1}{360(4\pi)^{2}}\left( N_{S}+\frac{11}{2}N_{F}+ 62
N_{V}\right) \label{TraceAnomaly3b} \,,
\end{eqnarray}
\end{subequations}
where $N_{S}$, $N_{F}$ and $N_{V}$ denote the number of fields of
spin 0, 1/2 and 1 respectively ($\hbar=1$). It turns out that the
coefficient $b''$ is regularisation dependent and is therefore not
considered to be part of the true conformal anomaly. We take this
into account by allowing $b''$ to vary.

Let us clearly state the assumptions we use to solve for the
dynamics that is driven by the conformal anomaly. Firstly, we
specialise to flat Friedmann-Lema\^itre-Robertson-Walker or FLRW
spacetimes in which the metric is given by $g_{\alpha\beta}=
\mathrm{diag} \left(-1,a^{2}(t),a^{2}(t),a^{2}(t)\right)$ where
$a(t)$ is the scale factor of the Universe in cosmic time $t$. As
usual, $H=\dot{a}/a$ is the Hubble parameter. We consider a
Universe in the presence of a) a non-zero cosmological constant,
b) the trace anomaly as a contribution to the quantum
stress-energy tensor and c) classical matter with constant
equation of state $\rho_{\mathrm{M}}=\omega p_{\mathrm{M}}$, where
$\omega
> -1$. Thirdly, we use covariant stress-energy conservation and
assume a perfect fluid form for the quantum density
$\rho_{\mathrm{Q}}$ and quantum pressure $p_{\mathrm{Q}}$
contributing to the trace anomaly:
\begin{equation}\label{perfectfluidform}
T_{\phantom{\mu}\nu,\mathrm{Q}}^{\mu} =( -\rho_{\mathrm{Q}},
p_{\mathrm{Q}},p_{\mathrm{Q}}, p_{\mathrm{Q}})\,.
\end{equation}
The relevant differential equation governing the dynamics driven
by the trace anomaly, derived from the Einstein field equations is
\cite{Koksma:2008jn}:
\begin{equation}\label{EFEGeneral}
9(1+\omega) H^{2}(t)+6\dot{H}(t)-3(1+\omega)\Lambda = - 8\pi
G\left[T_{\mathrm{Q}}+ (1-3\omega)\rho_{\mathrm{Q}} \right]\,.
\end{equation}
Here, the trace anomaly in FLRW spacetimes reads:
\begin{equation}\label{TraceAnomaly4}
T_{\mathrm{Q}}= 4b' \left\{ \dddot{H} +7\ddot{H}H +4\dot{H}^{2} +
18\dot{H}H^{2} +6H^{4} \right\} -6b''\left\{ \dddot{H} +7\ddot{H}H
+4\dot{H}^{2} + 12\dot{H}H^{2}\right\}   \,.
\end{equation}
The quantum density contributing to (\ref{EFEGeneral}) is given by
(also see \cite{Fischetti:1979ue, Hawking:2000bb} for details):
\begin{equation}\label{quantumdensity3}
\rho_{\mathrm{Q}}= 2 b'\left[
-2\ddot{H}H+\dot{H}^{2}-6\dot{H}H^{2} -3H^{4}\right] +3b''\left[
2\ddot{H}H -\dot{H}^{2}+6\dot{H}H^{2}\right]\,.
\end{equation}
To capture the leading order dynamics we work in quasi de Sitter
spacetime where $H(t)$ depends only mildly on time:
\begin{equation}\label{QdeSitterHubble}
\epsilon \equiv -\frac{\dot{H}}{H^{2}} = \mathrm{constant} \ll 1
\,,
\end{equation}
i.e.: we assume that $\epsilon$ is both small and time
independent. This truncates the trace anomaly
(\ref{TraceAnomaly4}) up to terms linear in $\dot{H}$:
\begin{equation}\label{TraceAnomaly4b}
T_{\mathrm{Q}}= 24 b' \left\{ 3\dot{H}H^{2} +H^{4} \right\} -72b''
\dot{H}H^{2}  \,.
\end{equation}
The expression for the quantum density (\ref{quantumdensity3})
reduces to:
\begin{equation}\label{quantumdensity4}
\rho_{\mathrm{Q}}= -6 b'\left[ 2\dot{H}H^{2} +H^{4}\right] +18 b''
\dot{H}H^{2}\,.
\end{equation}
We will examine both the exact forms (\ref{TraceAnomaly4}) and
(\ref{quantumdensity3}) in general FLRW spacetimes and their
truncated versions (\ref{TraceAnomaly4b}) and
(\ref{quantumdensity4}) in quasi de Sitter spacetime. Recall that
we should allow the $b''$ parameter to vary. Note that when
$b''=2b'/3$ the truncated versions are exact.

Truncating the expression for the trace anomaly is motivated by
the following realisation. Generally, higher derivative
contributions in an equation of motion have the tendency to
destabilise a system unless the initial conditions are highly
fine-tuned. Formally, this is known as the theorem of Ostrogradsky
and its relevance to Cosmology is outlined, for example, in
\cite{Woodard:2006nt}.

In the literature (see e.g. \cite{Pelinson:2003gn}), one only has
considered a Universe with radiation, a cosmological term and the
trace anomaly (radiation does not contribute classically to the
trace of the Einstein field equations). We incorporate matter with
constant equation of state parameter $\omega$.

Independently on whether one truncates the expressions for the
anomalous trace or quantum density, one can straightforwardly
solve for the asymptotes of (\ref{EFEGeneral}):
\begin{equation}\label{Asymptote1}
\left(H_{0}^{\mathrm{C,A}}\right)^2=\frac{-1 \pm \sqrt{1+64\pi G
b'\Lambda/3}}{32\pi G b'} \,.
\end{equation}
Here, $H_{0}^{\mathrm{C}}$ turns out to be the classical de Sitter
attractor, whereas $H_{0}^{\mathrm{A}}$ is a new, quantum anomaly
driven attractor. Note from equation (\ref{TraceAnomaly3b}) that
$b'<0$. We can write the above expression in a somewhat more
convenient form by defining the dimensionless parameter $\lambda$:
\begin{equation}\label{ScaleDependenceLambda}
\lambda=\frac{G\Lambda}{3} \,,
\end{equation}
that sets the scale for the cosmological constant $\Lambda$. We
expand (\ref{Asymptote1}) as $\lambda \ll 1$,
finding\footnote{Note the nomenclature in the literature is
somewhat misleading. Rather than calling (\ref{Asymptote2a}) the
classical de Sitter attractor, it would be more natural to denote
it with the quantum corrected classical attractor. Hence, the
quantum attractor (\ref{Asymptote2b}) should preferably be denoted
by anomaly driven attractor or Planck scale attractor. We will
nevertheless adopt the nomenclature existing in the literature.}:
\begin{subequations}\label{Asymptote2}
\begin{eqnarray}
H_{0}^{\mathrm{C}} &=& \sqrt{\frac{\Lambda}{3}}\left[1-8\pi
b'\lambda \right]
\label{Asymptote2a}\\
H_{0}^{\mathrm{A}} &=& \sqrt{\frac{-1}{16\pi G \, b'}
-\frac{\Lambda}{3}} \label{Asymptote2b}\,.
\end{eqnarray}
\end{subequations}
In the absence of a cosmological constant, the trace anomaly can
thus provide us with an inflationary scenario which has already
been appreciated by \cite{Starobinsky:1980te, Hawking:2000bb,
Shapiro:2003gm}. Finally, note these asymptotes are independent of
$b''$.

\section{The Trace Anomaly in Quasi de Sitter spacetime}

In this section we work in quasi de Sitter spacetime, where we
treat $\epsilon$ as a small and time independent constant which
allows us to neglect higher order derivative contributions,
motivated by the theorem of Ostrogradsky \cite{Woodard:2006nt}.

We have to distinguish two cases separately. In the spirit of
\cite{Antoniadis:2006wq}, the numerical value of the parameter
$b''$ occurring in the trace anomaly is not fixed and we therefore
allow it to take different values. First, we consider an
unrestricted value of $b''$, but where $b''\neq 2b'/3$, and
secondly we set $b''=2 b'/3$.

\subsection{Case I: unrestricted value of $\mathbf{b''}$}

\begin{figure}
    \begin{minipage}[t]{.48\textwidth}
        \begin{center}
\includegraphics[width=\textwidth]{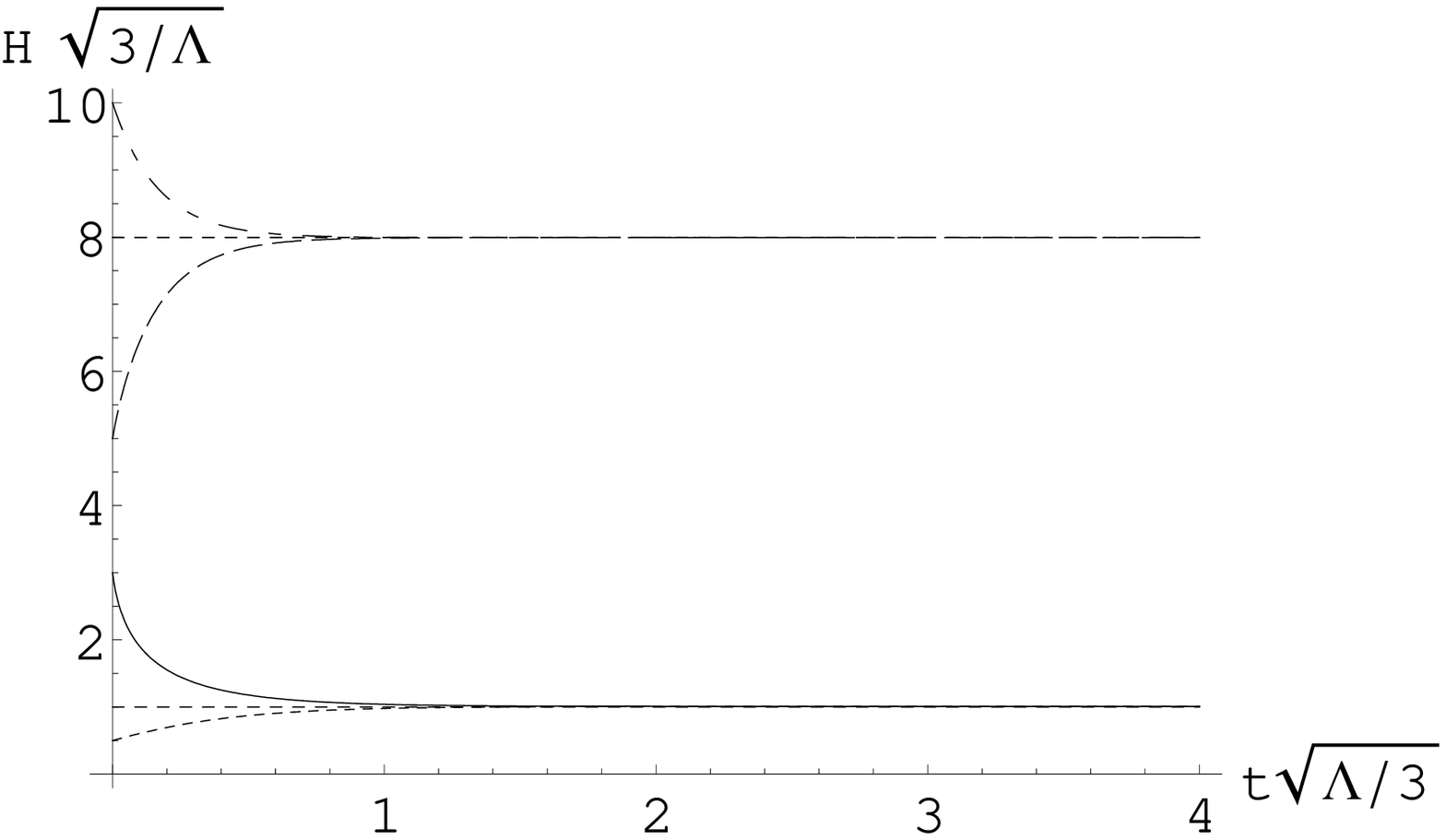}
        \end{center}
   \end{minipage}
\hfill
    \begin{minipage}[t]{.48\textwidth}
        \begin{center}
\includegraphics[width=\textwidth]{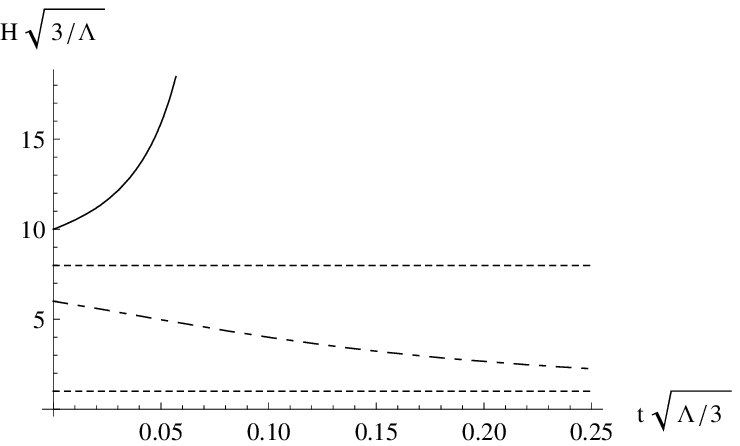}
        \end{center}
    \end{minipage}
{\em \caption{
   \textit{Left plot:} dynamics of the Hubble parameter in quasi de Sitter
   spacetime with a non-zero cosmological constant,
   the trace anomaly and matter $\omega=0$. Depending on the
   initial conditions, $H(t)$ evolves to either the
   classical de Sitter or the quantum anomaly driven attractor. We
   have used $\lambda=1/50$, $b'=-0.015$ (Standard Model
   value) and $b''=0$.
   \newline
   \textit{Right plot:} instability of the quantum anomaly driven attractor in quasi
   de Sitter spacetimes. We have used $\lambda=1/50$, $\omega=1/3$,
   $b''=7b'/6 < 5 b'/6$.
   \label{fig:QdS} }}
\end{figure}

We thus insert the truncated expression for the trace anomaly
(\ref{TraceAnomaly4b}) and the quantum density
(\ref{quantumdensity4}) into the Einstein field equation
(\ref{EFEGeneral}). This differential equation can be solved
exactly:
\begin{eqnarray}\label{Hsolution1}
t-t' &=& \frac{1}{-48\pi G b'(1+\omega)\left\{
(H_{0}^{\mathrm{A}})^{2}-(H_{0}^{\mathrm{C}})^{2} \right\}}
\left[- \frac{1+\frac{1}{2}\alpha
(H_{0}^{\mathrm{A}})^{2}}{H_{0}^{\mathrm{A}}}
\left\{\log\left(\frac{H(t)+H_{0}^{\mathrm{A}}}{H(t)-
H_{0}^{\mathrm{A}}}\right) -
\log\left(\frac{H(t')+H_{0}^{\mathrm{A}}}{H(t')-
H_{0}^{\mathrm{A}}}\right)\right\} \right. \\
 && \qquad\qquad\qquad\qquad\qquad \qquad\qquad\quad
 + \left. \frac{1+\frac{1}{2}\alpha
(H_{0}^{\mathrm{C}})^{2}}{H_{0}^{\mathrm{C}}}
\left\{\log\left(\frac{H(t)+H_{0}^{\mathrm{C}}}{H(t)-
H_{0}^{\mathrm{C}}}\right) -
\log\left(\frac{H(t')+H_{0}^{\mathrm{C}}}{H(t')-
H_{0}^{\mathrm{C}}}\right)\right\} \right]\nonumber  \,.
\end{eqnarray}
In the left plot of figure \ref{fig:QdS}, we numerically calculate
the dynamics of the Hubble parameter for various initial
conditions. The two asymptotes divide this graph into three
distinct regions that are not connected for finite time evolution.
The region bounded by the two asymptotes contains initial
conditions for $H(t)$ such that $H(t)$ grows for late times
towards $H_{0}^{\mathrm{A}}$ and initial conditions such that
$H(t)$ asymptotes to the de Sitter attractor $H_{0}^{\mathrm{C}}$,
separated by a branching point \cite{Koksma:2008jn}:
\begin{equation}\label{BranchingPoint}
H_{\mathrm{BP}}= \frac{1}{\sqrt{8\pi
G\left\{9(1+\omega)b''-2(5+3\omega)b'\right\}}} \,.
\end{equation}
Using the analytical solution (\ref{Hsolution1}), we can study the
stability of the late time attractors. The quantum anomaly driven
attractor is stable, whenever the following inequality is
satisfied:
\begin{equation}\label{StabilityQdSCond}
b'' > \frac{2}{9}b' \frac{4+3 \omega+16\pi\lambda b' (5+3\omega)}{
(1+\omega)(1+16\pi\lambda b') }\,.
\end{equation}
In the right plot of figure \ref{fig:QdS}, we numerically
calculate the evolution of the Hubble parameter in a radiation
dominated Universe when this inequality is not satisfied. We used
$b''=7b'/6 < 5 b'/6$. For initial conditions above
$H_{0}^{\mathrm{A}}$, the Hubble parameter increases to even
higher energies, whereas for initial conditions below the quantum
attractor, the Hubble parameter evolves towards the classical
attractor. The low energy limit reveals less surprising behaviour:
the classical de Sitter attractor is always stable.

\subsection{Case II: $\mathbf{b''}=2 \mathbf{b'}/3$}

As indicated earlier, we must consider the case when $b''=2 b'/3$
separately because in this particular case the total coefficient
in front of the $\Box R$ contribution to the trace anomaly
vanishes. All higher derivative contributions precisely cancel and
also the $\dot{H}^{2}$ contribution happens to cancel, such that
we find ourselves immediately situated in quasi de Sitter
spacetime. Albeit a simple case, we do take the full trace anomaly
into account.

The analytic solution obtained in (\ref{Hsolution1}) still applies
and moreover, it becomes exact. Most important, the features of
the left plot of figure \ref{fig:QdS}, e.g.: two stable attractors
and the occurrence of a branching point, do not change.

\section{The Trace Anomaly in FLRW Spacetimes}

\begin{figure}[t]
    \begin{minipage}[t]{.48\textwidth}
        \begin{center}
\includegraphics[width=\textwidth]{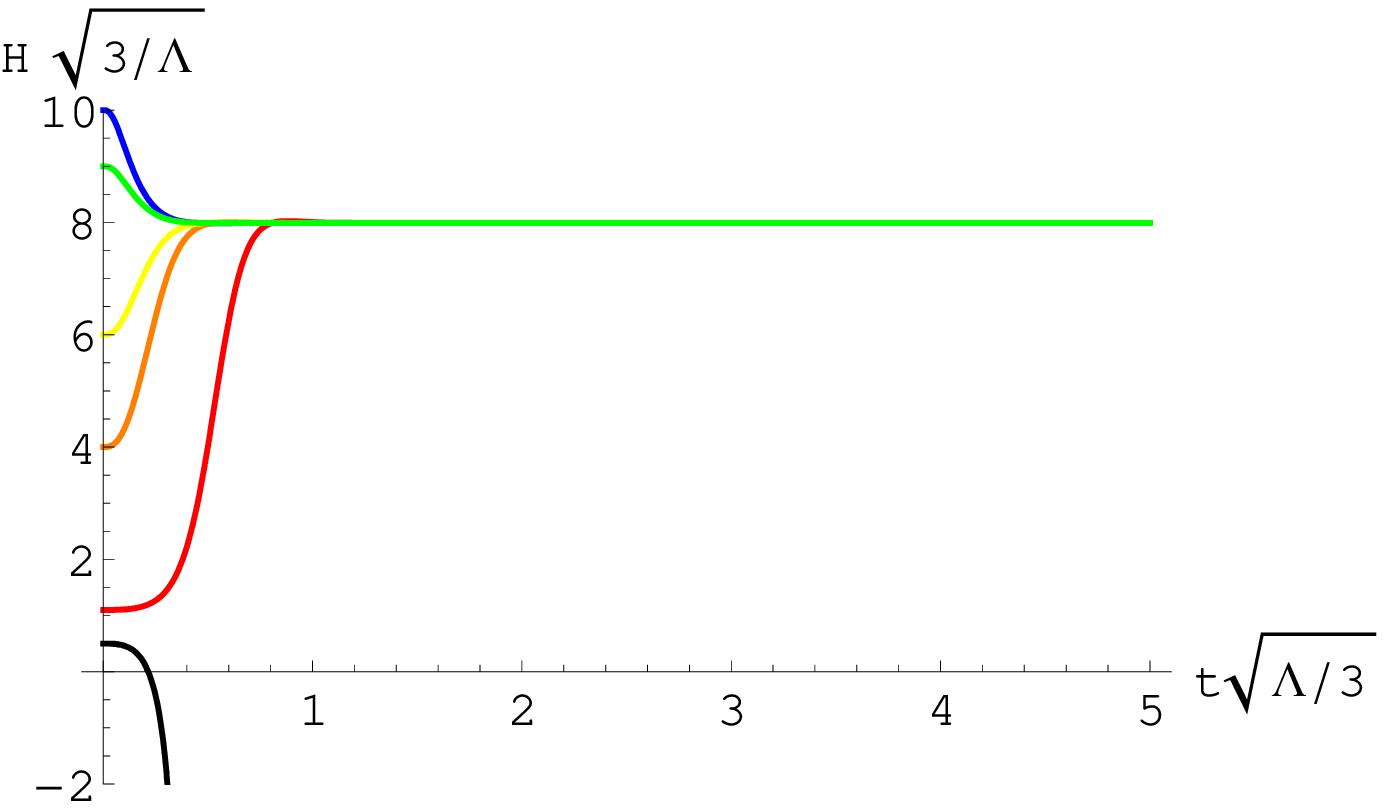}
         \end{center}
    \end{minipage}
\hfill
    \begin{minipage}[t]{.48\textwidth}
        \begin{center}
\includegraphics[width=\textwidth]{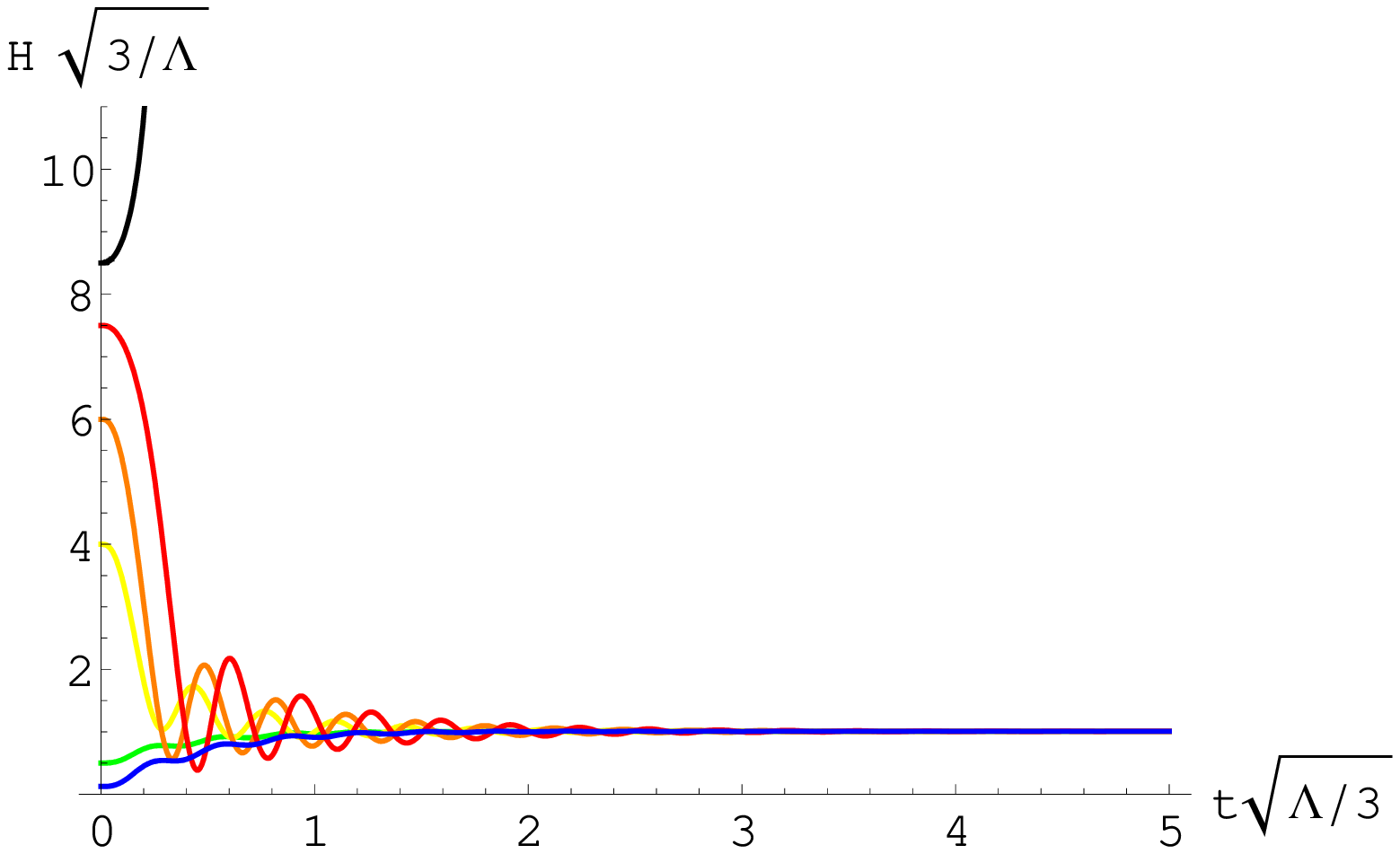}
        \end{center}
    \end{minipage}
{\em \caption{
   \textit{Left plot:} dynamics of the Hubble parameter taking the full trace anomaly
   into account. We took $b''=0$ such that $b''-2b'/3 > 0$
   yielding an unstable classical attractor. We have used $\omega=0$,
   $\lambda=1/50$, $b'=-0.015$ (Standard Model value).
   \newline
   \textit{Right plot:} dynamics of the Hubble parameter taking the full trace anomaly
   into account. We took $b''=b'$ such that $b''-2b'/3 < 0$
   yielding a stable classical attractor (the other parameters have not changed). Clearly, the classical attractor is under-damped, resulting in various
   oscillations around $H_{0}^{\mathrm{C}}$.
   \label{fig:FLRW} }}
\end{figure}

We turn our attention to solving the full trace equation
(\ref{EFEGeneral}), where we truncate the expression neither for
the anomalous trace (\ref{TraceAnomaly4}) nor for the quantum
density (\ref{quantumdensity3}). Before turning to numerical
methods, let us summarise what we can learn from analytically
studying the stability behaviour of the late time attractors. Let
us consider small perturbations $\delta H(t)$ around the two
asymptotes and insert:
\begin{equation}\label{PerturbationAsymptote}
H(t)= H_{0}^{\mathrm{C,A}} + \delta H(t)\,,
\end{equation}
in equation (\ref{EFEGeneral}), where $H_{0}^{\mathrm{C,A}}$ can
either denote the classical or the quantum attractor. If we make
the ansatz $\delta H(t)=c\exp[\xi t]$, we can solve the linearised
characteristic equation for the eigenvalues $\xi$. Clearly,
$\mathrm{Re}(\xi)<0$ ensures the stability of an attractor. This
yields:
\begin{subequations}
\begin{eqnarray}
\mathrm{If} \,\, b''-2b'/3 > 0, && \mathrm{then}\,\, \left\{
\begin{array}{l}
\mbox{Classical attractor unstable} \\
\mbox{Quantum attractor stable}
\end{array} \right.
\label{StabilityCondition1}\\
\mathrm{If} \,\, b''-2b'/3 < 0, && \mathrm{then}\,\, \left\{
\begin{array}{l}
\mbox{Classical attractor stable} \\
\mbox{Quantum attractor unstable}
\end{array} \right.
\label{StabilityCondition2}
\end{eqnarray}
\end{subequations}
Surprisingly, the stability analysis does not depend on the
equation of state $\omega$. This calculation thus proves the
statements about stability made in e.g. \cite{Pelinson:2002ef}
using the Routh-Hurwitz method. Our proof is more general because
we include a constant, but otherwise arbitrary, equation of state
parameter $\omega>-1$. Moreover, while the Routh-Hurwitz method
can only guarantee stability of a solution (when certain
determinants are all strictly positive), it does not tell anything
about instability \cite{Pelinson:2002ef}. Furthermore appreciate
that the singular point in this analysis, $b''-2b'/3=0$,
immediately directs us to the quasi de Sitter spacetime analysis
performed in the previous section, where all higher derivative
contributions precisely cancel rendering both attractors stable.

Let us compare the two plots in figure \ref{fig:FLRW}. In the left
plot, we used $b''=0$ such that $b''-2b'/3 > 0$, yielding an
unstable classical attractor. However, if $H(0) \leq
H_{0}^{\mathrm{C}}$ the quantum anomaly driven asymptote is not an
attractor and the Hubble parameter runs away to negative infinity.
In the right plot, we set $b''=b'$ such that $b''-2b'/3 < 0$ which
gives us a stable classical attractor. Likewise, for initial
conditions $H(0) \geq H_{0}^{\mathrm{A}}$ the de Sitter solution
is not an attractor and the Hubble parameter rapidly blows up to
positive infinity.

The right plot reveals another interesting phenomenon. In this
case, the classical attractor is under-damped, resulting in
decaying oscillations around this attractor. Clearly, oscillatory
behaviour occurs whenever the eigenvalues $\xi$ develop an
imaginary contribution. We can derive that whenever the classical
de Sitter attractor is stable oscillations occur. Oscillatory
behaviour around the quantum attractor occurs only when:
\begin{equation}\label{CondOscill5}
b'' < - \frac{2}{9} b'\left( \frac{1+8\pi \lambda b'}{1 + 8 \pi
\lambda}\right)\,.
\end{equation}

\section{Conclusion}

We have studied the dynamics of the Hubble parameter both in quasi
de Sitter and in FLRW spacetimes including matter, a cosmological
term and the trace anomaly. We have seen that there is no
dynamical effect that influences the effective value of the
cosmological constant, i.e.: the classical de Sitter attractor.
Based on our semiclassical analysis we thus conclude that the
trace anomaly does \textit{not} solve the cosmological constant
problem.

Ostrogradsky's theorem merits another remark. Clearly, including
the higher derivative contributions in FLRW spacetimes modifies
the dynamics of the Hubble parameter significantly: attractors,
that were stable in the absence of higher derivatives, under
certain conditions destabilise. We do not know which of the two
approaches is correct. Discarding these higher derivatives and
studying the trace anomaly in quasi de Sitter spacetime would seem
plausible.

Finally, one could wonder whether the quantum anomaly driven
attractor is physical. The quantum attractor is of the order of
the Planck mass $M_{\mathrm{pl}}$, so only when the matter in the
early Universe is sufficiently dense, $H \simeq \mathcal{O}(
M_{\mathrm{pl}})$. We then expect to evolve towards the quantum
attractor. However, at these early times we also expect
perturbative general relativity to break down. Hence, this
attractor may be seriously affected by quantum fluctuations or it
might even not be there.

\end{document}